\documentclass{article}

\usepackage{arxiv}
\usepackage[utf8]{inputenc} 
\usepackage[T1]{fontenc}    
\usepackage{hyperref}       
\usepackage{url}            
\usepackage{booktabs}       
\usepackage{amsfonts}       
\usepackage{nicefrac}       
\usepackage{microtype}      
\usepackage{lipsum}		
\usepackage{graphicx}
\usepackage[numbers]{natbib}
\usepackage{doi}
\usepackage{comment}

\title{An Analysis of Socialbots Activities and Influence in Modern Japanese Social Media}

\author{{Shuhei Ippa} \\
	Graduate School of Information Security\\
	Institute of Information Security\\
	Yokohama, Japan\\
	\texttt{mgs224502@iisec.ac.jp} \\
	\And
	\href{https://orcid.org/0000-0001-5596-282X}{\includegraphics[scale=0.06]{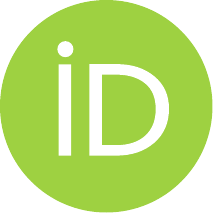}\hspace{1mm}Masaki Hashimoto} \\
	Graduate School of Information Security\\
	Institute of Information Security\\
	Yokohama, Japan\\
	\texttt{hashimoto@iisec.ac.jp} \\
}

\renewcommand{\shorttitle}{\textit{arXiv} Template}

\hypersetup{
pdftitle={A template for the arxiv style},
pdfsubject={q-bio.NC, q-bio.QM},
pdfauthor={Shuhei Ippa, Masaki Hashimoto},
pdfkeywords={First keyword, Second keyword, More},
}

\begin{document}
\maketitle

\begin{abstract}
In recent years, the proliferation of disinformation has become an issue against the backdrop of the spread of social media. In this study, we focus on socialbots, one of the causes of this problem, and analyze several domestic cases to clarify the actual activities and influence of socialbots. As a result of this analysis, we found that the influence of socialbots is greater in Japan than in the U.S. presidential election of 2016, which is a representative case of socialbot influence, and that socialbots retweeted by humans are not significantly different from human accounts. In addition, socialbot accounts retweeted by humans are not significantly different from human accounts. This paper also discusses specific methods and perspectives for further analysis and research on the influence of socialbots.
\end{abstract}


\section{Introduction}
\subsection{Background}
In today's Japan, the amount of time spent on the Internet is on the rise, with the percentage of people who use the Internet to quickly learn about events and movements in the world at 60\% overall and more than 70\% of those in their teens to 30s, as well as the use of social media such as LINE, Instagram, and Twitter, all of which are on the rise\cite{r4research}. The use of social media such as LINE, Instagram, and Twitter is also on the rise. Against this backdrop, disinformation on the Internet can be created and published more quickly and cheaply than traditional media such as newspapers and television, making it easier to attract public attention and making countermeasures against this issue a challenge\cite{shu2017fake}. Disinformation spreads significantly further, faster, deeper, and wider than other genres when it is related to politics, conspiracy theories, and business\cite{vosoughi2018spread}.

One of the factors contributing to the spread of disinformation on social media is the use of socialbots\cite{gorwa2020unpacking} that automatically tweet and retweet. Socialbots are accounts that send out useful information related to vaccination, earthquakes, etc.\cite{social_bots}, but some of them are used for the purpose of manipulating public opinion on social media. In fact, one study detected socialbots in 10,000 random accounts that tweeted news rated as fake, and found that socialbots accounted for 22\% of the tweets\cite{shu2020fakenewsnet}. Early socialbots were detected by a simple detection method that focused on a large number of posts because they were engaged in a single type of activity: automatically posting content\cite{ferrara2016rise}. However, the behavior of socialbots has become more sophisticated in recent years\cite{lee2011seven}.

\subsection{Purpose and Contribution of this Study}
The purpose of this study is to analyze the actual activities and influence of socialbots in the diffusion of information on social media in Japan today. In other words, we will take up several specific cases of socialbots used for the purpose of disinformation diffusion using existing tools, and compare the number and ratio of socialbots with the results of similar analyses in other countries. We compare the number and ratio of socialbots with similar analysis results in other countries, and clarify what kind of socialbots influence information diffusion.

The contribution of this study is to demonstrate the following points about modern Japanese social media by conducting the analysis in accordance with the above objectives.
\begin{itemize}
	\item Socialbots are more active than typical cases in English-speaking countries.
	\item Regardless of the content of the tweets, humans are spreading socialbots' posts.
	\item Socialbot accounts are not significantly different from human accounts.
\end{itemize}

\subsection{Structure of this paper}
In Chapter 2, we survey socialbot detection services and the latest research on socialbots in the United States. Then, we summarize the issues and present the position of this study. In Chapter 3, we present the case studies and the specific collection and analysis methods used in this study. In Chapter 4, we comprehensively present the results of the analysis in this study, and evaluate and discuss the numerical values of the percentage of socialbots. In Chapter 5, based on the evaluation and discussion in Chapter 4, we summarize this paper by presenting the methods and perspectives to be taken up for further analysis of the actual activities and influence of socialbots as future issues.

\section{Related Research}
\subsection{Socialbot Detection Tools}
In the detection of socialbots, Tabata et al\cite{Tabata2020twitter}. found that there are few research tools for Japanese-language accounts, but Zannettou et al\cite{zannettou2019disinformation}. Therefore, it is desirable to use a tool that is not specific to a particular language and can be used universally.

Botometer\cite{davis2016botornot} is an online tool developed by Indiana University in the U.S. After the release of v1 in May 2014, v2 was released in May 2016, v3 in May 2018, and the latest version, v4, was released in September 2020, receiving more than 500,000 socialbot decision requests per day\cite{sayyadiharikandeh2020detection}. Botometer v4 uses a dataset of approximately 64,000 accounts identified as socialbots (15,000 in v1) and 51,000 human accounts (16,000 in v1), and classifies them into six classes of 1,200+ features, ranging from 0 to 1. The bot score is calculated from 0 to 1 based on more than 1,200 features classified into 6 classes.

Botometer is considered a trusted tool for socialbot detection because it has been used in many studies, including a disinformation analysis of the novel coronavirus\cite{shahi2021exploratory}\cite{vosoughi2018spread}\cite{shu2020fakenewsnet}\cite{badawy2018analyzing}\cite{bessi2016social}\cite{shao2018spread}\cite{shao2018anatomy}\cite{varol2017online}\cite{broniatowski2018weaponized}.

Users of Botometer are referred to the English score, which includes features based on natural language processing, when targeting English-speaking accounts, and the universal score, which does not depend on the same features, when targeting non-English-speaking accounts\cite{yang2019arming}. Sugimori et al\cite{Sugimori2018}. have developed a universal language-independent classification model based on Botometer. The model is considered to be equivalent to the universal score of Botometer, and it was confirmed that approximately 95\% of the accounts could be correctly classified. This means that the accuracy of the universal score is high even for the language-independent Botometer universal score.

\subsection{Case Studies}
\begin{figure}[htb]
\begin{center}
\includegraphics[bb=0 0 261 195, scale=1.0]{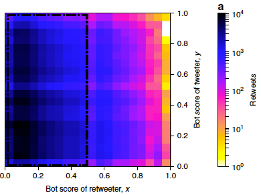}
\caption{Relationship between tweet accounts and retweet accounts}
\label{fig:heatmap_us}
\end{center}
\end{figure}

In the English-speaking world, between 9\% and 15\% of accounts by population are socialbots\cite{varol2017online}, and the 2016 U.S. presidential election was credited with enabling national and foreign governments to deploy legions of socialbots to influence the direction of the online conversation\cite{badawy2018analyzing}. Bessi et al. collected more than 20 million tweets by about 2.8 million users and estimated that about 400,000 socialbots, or one-fifth of the total, were responsible for about 3.8 million tweets\cite{bessi2016social}, while Shao et al. found that humans do not distinguish between humans and socialbots in their retweets, as shown in Figure\ref{fig:heatmap_us}\cite{shao2018spread}. In this regard, in another study, Shao et al. found that in the center of the retweet network for misinformation, 25\% of the retweets were human retweets of socialbots\cite{shao2018anatomy}. A similar phenomenon was observed in the 2016 referendum on whether the United Kingdom should leave the European Union\cite{gorodnichenko2021social}.

Other studies on socialbots include the Disinformation campaign called MacronLeaks in the 2017 French presidential election, where socialbots created shortly before the 2016 U.S. presidential election and used for only one week were used again\cite{ferrara2017disinformation}, the the 2016 UK referendum to leave the EU, that socialbots caused small to medium-scale disinformation to spread faster than humans\cite{bastos2019brexit}, and that the most active accounts for each side of leave and remain were socialbots and were used to retweet\cite{howard2016bots}.

The study on socialbots in Japan was conducted by Mintalra et al. during the 48th general election of the House of Representatives in 2017, with data collection conducted from October 10 to October 23, 2017, where socialbots accounted for about 2\% (2,411 out of 94,451 accounts ), and socialbots tweets were about 4\% of all relevant tweets (28,531 out of 665,400 tweets, including an additional 13,404 tweets from deleted accounts), which is different from the United States, the United Kingdom, and France, where socialbots tweets range from 12.5\% to 25\%\cite{mintal2019trendy}. On the other hand, Mintalra et al. state that the role of socialbots during political events has been studied in various studies published in Western democracies, but has rarely been studied outside Western countries such as Japan. Incidentally, Botometer was not used for socialbots determination in this study.

\subsection{Summary of related research}
Overseas, research has been conducted on socialbots, including the development of Botometer, which estimates that between 9\% and 15\% of accounts in the English-speaking world are socialbots by population, that socialbots accounted for about 1/7th of accounts in the 2016 U.S. presidential election, that socialbots tweeted about 1/5th of the tweets, and that humans do not distinguish between humans and socialbots in retweeting. In the 2016 U.S. presidential election, it was estimated that socialbots, which accounted for about 1/7th of the accounts, tweeted about 1/5 of the tweets, and that humans do not distinguish between humans and socialbots in retweets.

This study is a first step toward analyzing socialbot activity and impact in Japan, using the Botometer, which has been used in overseas studies, to determine the proportion of socialbots in Japan.

\section{Data Collection and Analysis Methodology}
\subsection{Data Collection}
While this study will focus on the activity and impact of socialbots, Disinformation will focus on the characteristics that Disinformation spreads significantly farther, faster, deeper, and wider when the topic is politics, conspiracy theories, or business-related, and will take up examples.

In this study, we first take up a topic that has been the subject of much controversy in recent years: the state funeral of former Prime Minister Abe. The state funeral of former Prime Minister Abe were reportedly tweeted using socialbots\cite{yomiuri}, and the fact that there was a sharp division of opinion on the implementation of the rites makes us believe that this is a case in which the activities of socialbots are likely to promote polarization. We collected Twitter data on tweets about the word "state funeral" from July 8, 2022 to September 27, 2022, the period between the date of the shooting of former Prime Minister Abe and the date of the state funeral, via the Twitter API. The data was obtained from 980,894 accounts with 21,343, 117 tweets.

Next, as a case study in the political field, where disinformation is the most far-reaching, fast, deep, and wide-spread topic, we will discuss the most recent political event, the unified local elections held in April 2023. Unified local elections are nationwide elections for the heads of local governments and members of local assemblies, with the purpose of raising voters' awareness of elections nationwide. We judged that the election was the most notable large-scale political event in the absence of national elections such as the general election for the House of Representatives and the regular election for the House of Councillors. Data collection was conducted for the term "unified local elections" from March 23, 2023 to April 23, 2023, the date of the election announcement to the date of the vote, yielding 440,601 data items from 143,440 accounts.

Lastly, we will discuss the discourse that Shikishima Baking Company is glorifying crickets for the sake of government subsidies, which was declared false by the Japan Fact Check Center\cite{factcheckcenter} in March 2023 (hereinafter referred to as the PASCO case). This case was selected not only because it was actually determined to be disinformation\cite{pasco}, but also because it is a case related to conspiracy theories and business, which are considered to be the most conspicuous sources of disinformation after politics. Data was collected using the keyword "PASCO cricket" from February 26, 2023 to March 10, 2023, the date when the Japan Fact-Checking Center announced the tweet as a disinformation, and 98,297 data were obtained from 49,467 accounts.

\subsection{Analysis Methodology}
Botometer is a tool targeting Twitter accounts, and the number of tweets on Twitter in disinformation cases is approximately 10 times the number of posts on Facebook and Instagram (17 times at peak times)\cite{ruck2019internet}. Since Twitter tends to be used for disinformation diffusion, this study focuses on Twitter accounts and tweets related to each case study. In this study, we adopt 0.5 as the threshold of the bot score to judge a social bot as a social bot, which has been used as a standard in many previous studies
\cite{vosoughi2018spread}
\cite{shu2020fakenewsnet}
\cite{badawy2018analyzing}
\cite{bessi2016social}
\cite{shao2018spread}
\cite{shao2018anatomy}
\cite{varol2017online}.

After analyzing the numerical values of the proportion of socialbots in each case, we compared the following 10 items displayed on the Twitter account page with those of human retweeters in order to understand the characteristics of socialbots when humans retweet socialbots, as was confirmed in the 2016 U.S. presidential election and the referendum on whether or not the U.K. should leave the European Union. This is an examination of what items are characteristically different between socialbots and humans with respect to the target accounts of retweeting. This validation is based on the fact that research on the ability of humans to recognize socialbots has been identified as a future challenge\cite{bessi2016social}.
\begin{itemize}
	\item Number of Tweets Per Day (estimated from total number of tweets)
	\item Profile Picture
	\item Background Image
	\item Certification
	\item Self-Introductory Statement
	\item Residence
	\item URL Information
	\item Duration of Account Activity (estimated from the date of account creation)
	\item Number of Followings
	\item Number of Followers
\end{itemize}

\section{Evaluation and Discussion}
\subsection{Evaluation(State Funeral)}
\begin{figure}[htb]
\begin{center}
\includegraphics[bb=0 0 440 330, scale=1.0]{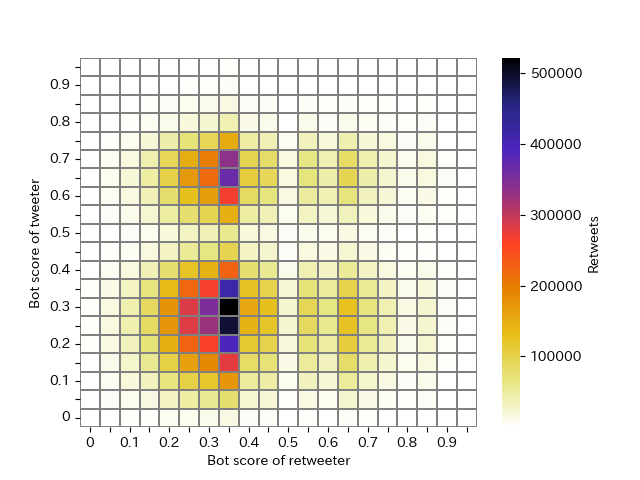}
\caption{Relationship between tweeting and retweeting accounts in united local elections}
\label{fig:k_h}
\end{center}
\end{figure}

We checked the number and percentage of socialbot accounts that tweeted about funerals, and found that 170,533 accounts, or 17.4 percent of the total 980,714 accounts, were socialbots. Socialbot tweets accounted for 6,414,556 out of 21,343,117 tweets, or 30.1 percent. The relationship between tweet accounts and retweet accounts was visualized using a heat map, as shown in Figure\ref{fig:k_h}, and it was confirmed that humans retweeted tweets from socialbots in retweeting.

The accounts that were retweeted (19,648 socialbot accounts and 112,312 human accounts) were organized as shown in table\ref{tab:k_account}, and it was found that the proportion of accounts with profile images and self-introductions was high for socialbots, and there was no significant difference between socialbot accounts and human accounts. The results show that there is no significant difference between socialbot accounts and human accounts. The absence of any authenticated accounts for both socialbot and human accounts may be due to the fact that Twitter had removed all authenticated accounts simultaneously in April 2023, the month in which the account analysis was conducted. Although there were 23 accounts with more than one million followers among the retweeted socialbots, most of these counts were related to major domestic and foreign media, and there were no accounts that were clearly considered socialbots.
\begin{table}[htb] 
\caption{Comparison of retweet destination accounts in state funeral}
\label{tab:k_account}
\hbox to\hsize{\hfil
\scalebox{1.7}[1.7]{
\scriptsize
\begin{tabular}{|c|rc|rc|} \hline
\multicolumn{1}{|c|}{Item} & \multicolumn{2}{c|}{Socialbot} & \multicolumn{2}{c|}{Human} \\\hline
Number of Tweets & 0-1 & 39.1\% & 0-1 & 24.6\% \\
Per Day & 2-5 & 18.3\% & 2-5 & 23.3\% \\
 & 6-10 & 10.3\% & 6-10 & 15.3\% \\
 & 11-20 & 10.3\% & 11-20 & 15.2\% \\
 & 21-50 & 11.6\% & 21-50 & 14.7\% \\
 & 51-100 & 6.5\% & 51-100 & 5.1\% \\
 & Over 100 & 4.3\% & Over 100 & 1.8\% \\
 \hline
Profile Picture & YES & 90.0\% & YES & 96.6\% \\
 & NO & 10.0\% & NO & 3.4\% \\
 \hline
Background Image & YES & 59.1\% & YES & 75.2\% \\
 & NO & 40.9\% & NO & 24.8\% \\
 \hline
Certification & YES & 0.0\% & YES & 0.0\% \\
 & NO & 100.0\% & NO & 100.0\% \\
 \hline
Self-Introductory Statement & YES & 80.7\% & YES & 90.9\% \\
 & NO & 19.3\% & NO & 9.1\% \\
 \hline
Residence & YES & 41.4\% & YES & 53.3\% \\
 & NO & 58.6\% & NO & 46.7\% \\
 \hline
URL Information & YES & 21.9\% & YES & 22.6\% \\
 & NO & 78.1\% & NO & 77.4\% \\
 \hline
Duration of & Within 1 week & 0.0\% & Within 1 week & 0.1\% \\
Account Activity & Within 1 month & 0.0\% & Within 1 month & 0.0\% \\
 & Within 1 year & 8.8\% & Within 1 year & 3.4\% \\
 & Over 1 year & 91.2\% & Over 1 year & 96.5\% \\
 \hline
Followings & 0 & 5.9\% & 0 & 0.5\% \\
 & 1-10 & 11.2\% & 1-10 & 1.6\% \\
 & 11-100 & 23.8\% & 11-100 & 14.2\% \\
 & 101-1,000 & 30.2\% & 101-1,000 & 55.9\% \\
 & 1,001-10,000 & 26.0\% & 1,001-10,000 & 27.3\% \\
 & Over 10,000 & 2.9\% & Over 10,000 & 0.5\% \\
 \hline
Followers & 0 & 5.0\% & 0 & 0.5\% \\
 & 1-10 & 14.9\% & 1-10 & 4.0\% \\
 & 11-100 & 21.3\% & 11-100 & 19.6\% \\
 & 101-1,000 & 25.8\% & 101-1,000 & 49.1\% \\
 & 1,001-10,000 & 25.9\% & 1,001-10,000 & 23.8\% \\
 & Over 10,000 & 7.1\% & Over 10,000 & 3.0\% \\
\hline
\end{tabular}
}\hfil}
\end{table}

\subsection{Evaluation(Unified Local Elections)}
\begin{figure}[htb]
\begin{center}
\includegraphics[bb=0 0 440 330, scale=1.0]{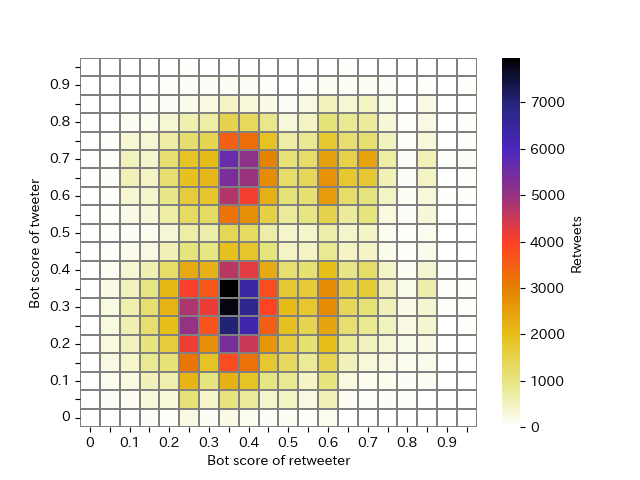}
\caption{Relationship between tweeting and retweeting accounts in united local elections}
\label{fig:t_h}
\end{center}
\end{figure}

We checked the number and percentage of socialbot accounts that tweeted about the nationwide local elections, and found that 34,125 accounts, or 23.8\% of the total 143,440 accounts, were socialbots. Socialbot tweets accounted for 164,915, or 37.4\%, of the 440,601 tweets in the nationwide local elections. The relationship between tweet accounts and retweet accounts was visualized using a heat map, as shown in the figure\ref{fig:t_h}, and it can be confirmed that humans retweeted the tweets of socialbots, similar to the results of the analysis on the national funeral.

The table\ref{tab:t_account} shows the number of retweeted accounts (2,161 socialbot accounts and 6,622 human accounts), and it is notable that more than 10\% of the socialbot accounts were authenticated, which is significantly higher than the 2.5\% of human accounts. Among the socialbots, there were 229 authenticated accounts, all of which were related to local governments, politicians, and mass media, and the same was true for accounts with more than one million followers.
\begin{table}[htb] 
\caption{Comparison of retweet destination accounts in united local elections}
\label{tab:t_account}
\hbox to\hsize{\hfil
\scalebox{1.7}[1.7]{
\scriptsize
\begin{tabular}{|c|rc|rc|} \hline
\multicolumn{1}{|c|}{Item} & \multicolumn{2}{c|}{Socialbot} & \multicolumn{2}{c|}{Human} \\\hline
Number of Tweets & 0-1 & 38.0\% & 0-1 & 23.8\% \\
Per Day & 2-5 & 19.5\% & 2-5 & 22.2\% \\
 & 6-10 & 9.6\% & 6-10 & 13.1\% \\
 & 11-20 & 8.8\% & 11-20 & 15.0\% \\
 & 21-50 & 10.7\% & 21-50 & 16.0\% \\
 & 51-100 & 7.6\% & 51-100 & 7.0\% \\
 & Over 100 & 5.8\% & Over 100 & 2.9\% \\
 \hline
Profile Picture & YES & 97.5\% & YES & 98.4\% \\
 & NO & 2.5\% & NO & 1.6\% \\
 \hline
Background Image & YES & 81.2\% & YES & 83.9\% \\
 & NO & 18.8\% & NO & 16.1\% \\
 \hline
Certification & YES & 10.6\% & YES & 2.5\% \\
 & NO & 89.4\% & NO & 97.5\% \\
 \hline
Self-Introductory Statement & YES & 93.1\% & YES & 95.0\% \\
 & NO & 6.9\% & NO & 5.0\% \\
 \hline
Residence & YES & 56.4\% & YES & 60.8\% \\
 & NO & 43.6\% & NO & 39.2\% \\
 \hline
URL Information & YES & 49.8\% & YES & 40.4\% \\
 & NO & 50.2\% & NO & 59.6\% \\
 \hline
Duration of & Within 1 week & 0.4\% & Within 1 week & 0.1\% \\
Account Activity & Within 1 month & 1.6\% & Within 1 month & 0.5\% \\
 & Within 1 year & 17.5\% & Within 1 year & 8.9\% \\
 & Over 1 year & 80.5\% & Over 1 year & 90.5\% \\
 \hline
Followings & 0 & 3.3\% & 0 & 0.3\% \\
 & 1-10 & 6.8\% & 1-10 & 0.7\% \\
 & 11-100 & 20.6\% & 11-100 & 9.7\% \\
 & 101-1,000 & 31.2\% & 101-1,000 & 52.0\% \\
 & 1,001-10,000 & 34.5\% & 1,001-10,000 & 35.8\% \\
 & Over 10,000 & 3.6\% & Over 10,000 & 1.5\% \\
 \hline
Followers & 0 & 0.5\% & 0 & 0.1\% \\
 & 1-10 & 3.2\% & 1-10 & 1.0\% \\
 & 11-100 & 11.0\% & 11-100 & 9.5\% \\
 & 101-1,000 & 30.7\% & 101-1,000 & 42.1\% \\
 & 1,001-10,000 & 41.2\% & 1,001-10,000 & 38.8\% \\
 & Over 10,000 & 13.4\% & Over 10,000 & 8.5\% \\
\hline
\end{tabular}
}\hfil}
\end{table}

\subsection{Evaluation(PASCO Case)}
\begin{figure}[htb]
\begin{center}
\includegraphics[bb=0 0 440 330, scale=1.0]{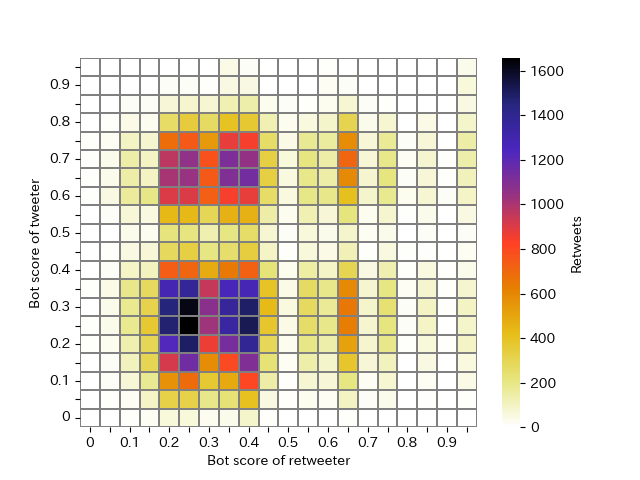}
\caption{Relationship between tweeting and retweeting accounts in PASCO case}
\label{fig:p_h}
\end{center}
\end{figure}

We checked the number and percentage of socialbot accounts that tweeted about the PASCO case study. 14,320 accounts, or 28.9\% of the total 49,467 accounts, were socialbot accounts. Socialbot tweets accounted for 36,494 out of 98,297 tweets, or 37.1\% of the total number of tweets in the PASCO case. The relationship between tweet accounts and retweet accounts was visualized using a heat map, as shown in the figure\ref{fig:p_h}, and the characteristics of humans retweeting socialbots were similar to those in the other cases.

The table shows that more than 10\% of the retweeted accounts (441 socialbot accounts and 2,094 human accounts) had more than 10,000 followers (4.0\% of human accounts), suggesting the existence of so-called "influencers". The difference from the other two cases is that no socialbot with more than one million followers was identified and only seven accounts (1.6\%) with more than 100,000 followers were found.
\begin{table}[htb] 
\caption{Comparison of retweet destination accounts in PASCO case}
\label{tab:p_account}
\hbox to\hsize{\hfil
\scalebox{1.7}[1.7]{
\scriptsize
\begin{tabular}{|c|rc|rc|} \hline
\multicolumn{1}{|c|}{Item} & \multicolumn{2}{c|}{Socialbot} & \multicolumn{2}{c|}{Human} \\\hline
Number of Tweets & 0-1 & 21.3\% & 0-1 & 19.3\% \\
Per Day & 2-5 & 11.8\% & 2-5 & 21.2\% \\
 & 6-10 & 13.2\% & 6-10 & 15.2\% \\
 & 11-20 & 13.2\% & 11-20 & 16.9\% \\
 & 21-50 & 18.6\% & 21-50 & 18.1\% \\
 & 51-100 & 12.9\% & 51-100 & 7.0\% \\
 & Over 100 & 9.0\% & Over 100 & 2.3\% \\
 \hline
Profile Picture & YES & 95.5\% & YES & 97.7\% \\
 & NO & 4.5\% & NO & 2.3\% \\
 \hline
Background Image & YES & 71.2\% & YES & 77.2\% \\
 & NO & 28.8\% & NO & 22.8\% \\
 \hline
Certification & YES & 0.2\% & YES & 0.2\% \\
 & NO & 99.8\% & NO & 99.8\% \\
 \hline
Self-Introductory Statement & YES & 85.7\% & YES & 92.9\% \\
 & NO & 14.3\% & NO & 7.1\% \\
 \hline
Residence & YES & 39.0\% & YES & 49.2\% \\
 & NO & 61.0\% & NO & 50.8\% \\
 \hline
URL Information & YES & 26.3\% & YES & 19.1\% \\
 & NO & 73.7\% & NO & 80.9\% \\
 \hline
Duration of & Within 1 week & 0.0\% & Within 1 week & 0.0\% \\
Account Activity & Within 1 month & 0.2\% & Within 1 month & 0.1\% \\
 & Within 1 year & 16.1\% & Within 1 year & 8.9\% \\
 & Over 1 year & 83.7\% & Over 1 year & 91.0\% \\
 \hline
Followings & 0 & 2.9\% & 0 & 0.5\% \\
 & 1-10 & 5.4\% & 1-10 & 1.3\% \\
 & 11-100 & 20.2\% & 11-100 & 13.1\% \\
 & 101-1,000 & 26.8\% & 101-1,000 & 53.0\% \\
 & 1,001-10,000 & 38.1\% & 1,001-10,000 & 30.6\% \\
 & Over 10,000 & 6.6\% & Over 10,000 & 1.5\% \\
 \hline
Followers & 0 & 2.3\% & 0 & 0.2\% \\
 & 1-10 & 7.7\% & 1-10 & 2.2\% \\
 & 11-100 & 15.4\% & 11-100 & 17.6\% \\
 & 101-1,000 & 28.6\% & 101-1,000 & 49.1\% \\
 & 1,001-10,000 & 35.4\% & 1,001-10,000 & 26.9\% \\
 & Over 10,000 & 10.6\% & Over 10,000 & 4.0\% \\
\hline
\end{tabular}
}\hfil}
\end{table}

\subsection{Discussion}
The number and percentage of socialbots and the number and percentage of tweets by socialbots in the three cases in this study and in the 2016 U.S. presidential election can be summarized as shown in the table.
\begin{table}[htb] 
\caption{Summary of socialbot percentages, etc.}
\label{tab:summary_table}
\hbox to\hsize{\hfil
\scalebox{1.0}[1.0]{
\scriptsize
\begin{tabular}{|c|c|c|c|c|} \hline
\multicolumn{1}{|c|}{} & 
\multicolumn{1}{c|}{State Funeral} & \multicolumn{1}{c|}{Unified Local Elections} & \multicolumn{1}{c|}{PASCO Case} & \multicolumn{1}{c|}{U.S. Presidential Election} \\\hline
Percentage of & 17.4\% & 23.8\% & 28.9\% & About 14\% \\
Socialbots & (170,533 / 980,714) & (34,125 / 143,440) & (14,320 / 49,467) & (About 400,000 / About 2.8 million) \\
\hline
Percentage of Tweets & 30.1\% & 37.4\% & 37.1\% & About 19\% \\
by Socialbots & (6,414,556 / 21,343,117) & (164,915 / 440,601) & (36,494 / 98,297) & (About 3.8 million / About 20 million) \\
\hline
\end{tabular}
}\hfil}
\end{table}
In the three cases analyzed in this study, the percentage of socialbots exceeded that of the 2016 U.S. presidential election. The same is true for the proportion of tweets by socialbots, which exceeds 30\%. The highest percentage of socialbots was found in the PASCO case, demonstrating that socialbots are a factor in the spread of disinformation.

Since it was confirmed that humans retweeted socialbots in three cases, the socialbots that were the target of retweets were organized as shown in the table\ref{tab:all_account}.
\begin{table}[htb] 
\caption{Comparison of retweet destination socialbots in each case study}
\label{tab:all_account}
\hbox to\hsize{\hfil
\scalebox{1.3}[1.3]{
\scriptsize
\begin{tabular}{|c|rc|rc|rc|} \hline
\multicolumn{1}{|c|}{Item} & 
\multicolumn{2}{c|}{State Funeral} & \multicolumn{2}{c|}{Unified Local Elections} & \multicolumn{2}{c|}{PASCO Case} \\\hline
Number of Tweets & 0-1 & 39.1\% & 0-1 & 38.0\% & 0-1 & 21.3\% \\
Per Day & 2-5 & 18.3\% & 2-5 & 19.5\% & 2-5 & 11.8\% \\
 & 6-10 & 10.3\% & 6-10 & 9.6\% & 6-10 & 13.2\% \\
 & 11-20 & 9.9\% & 11-20 & 8.8\% & 11-20 & 13.2\% \\
 & 21-50 & 11.6\% & 21-50 & 10.7\% & 21-50 & 18.6\% \\
 & 51-100 & 6.5\% & 51-100 & 7.6\% & 51-100 & 12.9\% \\
 & Over 100 & 4.3\% & Over 100 & 5.8\% & Over 100 & 9.0\% \\
 \hline
Profile Picture & YES & 90.0\% & YES & 97.5\% & YES & 95.5\% \\
 & NO & 10.0\% & NO & 2.5\% & NO & 4.5\% \\
 \hline
Background Image & YES & 59.1\% & YES & 81.2\% & YES & 71.2\% \\
 & NO & 40.9\% & NO & 18.8\% & NO & 28.8\% \\
 \hline
Certification  & YES & 0.0\% & YES & 10.6\% & YES & 0.2\% \\
 & NO & 100.0\% & NO & 89.4\% & NO & 99.8\% \\
 \hline
Self-Introductory Statement & YES & 80.7\% & YES & 93.1\% & YES & 85.7\% \\
 & NO & 19.3\% & NO & 6.9\% & NO & 14.3\% \\
 \hline
Residence & YES & 41.4\% & YES & 56.4\% & YES & 39.0\% \\
 & NO & 58.6\% & NO & 43.6\% & NO & 61.0\% \\
 \hline
URL Information & YES & 21.9\% & YES & 49.8\% & YES & 26.3\% \\
 & NO & 78.1\% & NO & 50.2\% & NO & 73.7\% \\
 \hline
Duration of & Within 1 week & 0.0\% & Within 1 week & 0.4\% & Within 1 week & 0.0\% \\
Account Activity & Within 1 month & 0.0\% & Within 1 month & 1.6\% & Within 1 month & 0.2\% \\
 & Within 1 year & 8.8\% & Within 1 year & 17.5\% & Within 1 year & 16.1\% \\
 & Over 1 year & 91.2\% & Over 1 year & 80.5\% & Over 1 year & 83.7\% \\
 \hline
Followings & 0 & 5.9\% & 0 & 3.3\% & 0 & 2.9\% \\
 & 1-10 & 11.2\% & 1-10 & 6.8\% & 1-10 & 5.4\% \\
 & 11-100 & 23.8\% & 11-100 & 20.6\% & 11-100 & 20.2\% \\
 & 101-1,000 & 30.2\% & 101-1,000 & 31.2\% & 101-1,000 & 26.8\% \\
 & 1,001-10,000 & 26.0\% & 1,001-10,000 & 34.5\% & 1,001-10,000 & 38.1\% \\
 & Over 10,000 & 2.9\% & Over 10,000 & 3.6\% & Over 10,000 & 6.6\% \\
 \hline
Followers & 0 & 5.0\% & 0 & 0.5\% & 0 & 2.3\% \\
 & 1-10 & 14.9\% & 1-10 & 3.2\% & 1-10 & 7.7\% \\
 & 11-100 & 21.3\% & 11-100 & 11.0\% & 11-100 & 15.4\% \\
 & 101-1,000 & 25.8\% & 101-1,000 & 30.7\% & 101-1,000 & 28.6\% \\
 & 1,001-10,000 & 25.9\% & 1,001-10,000 & 41.2\% & 1,001-10,000 & 35.4\% \\
 & Over 10,000 & 9.1\% & Over 10,000 & 13.4\% & Over 10,000 & 10.6\% \\
\hline
\end{tabular}
}\hfil}
\end{table}
In all cases, the number of tweets per day by the socialbot was less than one.
In all cases, the number of tweets per day by socialbots was less than one, and it was confirmed again that they did not engage in the activity of posting a large number of tweets, which was considered to be a characteristic of early socialbots.

In the Unified Local Elections, more than 10\% of the socialbots were authenticated, a significant difference from the other cases. These accounts are all related to local governments, politicians, and mass media, and are considered to be false positives that were determined to be social bots. As for the Botometer false positives, the problem of high bot scores for official accounts has already been pointed out\cite{rauchfleisch2020false}.

We know that two of the social bots with more than 100,000 followers in the PASCO case study include web media. Kawashima et al. found that the disinformation of politicians in Japan spread through intermediate media called "middle media"\cite{kawashima2019diffusion}. Kawashima et al. found that politicians' disinformation was spread via middle media in Japan, and it is possible that these two web media accounts were involved in the spread of misinformation in the PASCO case.

\section{Conclusion}
\subsection{Future Work}
In analyzing the PASCO case, which is a Disinformation case in this study, the existence of middle media that are pointed out to play a role in generating Disinformation became clear, and it can be inferred that these accounts were influential in the spread of information. On the other hand, the degree of influence exercised by these accounts has not been analyzed, and it is also possible that there were other influential accounts. In the future, we would like to define an index of influence in information diffusion and further identify accounts that exerted influence, including socialbots.

In addition, We also intend to conduct similar analysis of overseas cases.

\subsection{Summary}
This study analyzed the number and percentage of socialbots and retweets related to socialbots in several domestic cases against the background that socialbots sometimes work to spread disinformation and manipulate public opinion.

As a result, we found that the proportion of socialbots exceeded 15\% in all of the cases analyzed in this study, and that tweets by socialbots accounted for approximately 30\% of the tweets. This percentage exceeds the percentage of socialbots and tweets in the 2016 U.S. presidential election, where previous research on socialbots was conducted. In addition, humans retweeted the tweets of socialbots, and these socialbots did not differ significantly from human accounts in terms of profile images, self-introductions, and other settings, indicating the existence of socialbots with some influence.

\end{document}